\documentstyle[prl,aps,epsf]{revtex}
\draft
\def\ba{\begin{eqnarray}}
\def\ea{\end{eqnarray}}
\def\be{\begin{equation}}
\def\ee{\end{equation}}
\def\FX{F_{,X}}

\def\tmu{T_{,\mu}}
\def\phimu{\phi_{,\mu}}
\def\tnu{T_{,\nu}}
\def\phinu{\phi_{,\nu}}
\def\parX{F_{,X}}

\def\gmunu{g_{\mu\nu}}
\def\gupmunu{g^{\mu\nu}}

\begin{document}
%\magnification=\magstep1
\preprint{UPR-1000-T}
\twocolumn[\hsize\textwidth\columnwidth\hsize\csname
@twocolumnfalse\endcsname

\title{\large \bf Rolling Tachyon in Brane World Cosmology from 
Superstring Field Theory}
\medskip
\author{
Gary Shiu$^{1}$, S.-H.~Henry Tye$^2$, and Ira Wasserman$^{3}$ 
\vspace{0.2cm}}
\address{$^1$ Department of Physics and
Astronomy, University of Pennsylvania, Philadelphia, PA 19104 \\
$^2$ Laboratory for Elementary Particle Physics, Cornell University,
Ithaca, NY 14853 \\
$^3$ Center for Radiophysics and Space Research, Cornell University,
Ithaca, NY 14853}

\medskip
\date{\today}
\maketitle

{\tighten
\begin{abstract}
The pressureless tachyonic matter recently found in superstring field 
theory has an over-abundance problem in cosmology. We argue that 
this problem is naturally solved in the brane inflationary scenario 
if almost all of the tachyon energy is drained (via coupling to the 
inflaton and matter fields) to heating the Universe, 
while the rest of the tachyon 
energy goes to a network of cosmic strings (lower-dimensional BPS 
D-branes) produced during the tachyon rolling at the end of inflation. 
\end{abstract}
}%end tighten
]\narrowtext %% end two-column
%------%%%%%%%%%%%%%%%%%%%%%%%%%%%%%%%%%%%%%%%%%%%%%%%%%%%%%%%%%%%%%%%%

In superstring theory, starting with a non-BPS brane or a
brane-anti-brane pair, a tachyon is always present. This
tachyon rolls down its potential and tachyon
matter appears as energy density without pressure,
as pointed out by Sen\cite{sen1}.
In the cosmological context\cite{gibbons}, it is easy to estimate 
that the pressureless tachyonic matter density is
generically many orders of magnitude too big to be compatible 
with present day cosmological observations\cite{SW}.
This implies either that tachyon rolling does not happen 
(i.e., irrelevant) in the early Universe after inflation, or, 
instead of pressureless tachyon matter density, 
the tachyon potential energy goes somewhere else.
In this paper, we argue for the latter.

In superstring theory, defects, 
or stable solitons (i.e., lower-dimensional BPS branes),
are produced during tachyon rolling,
also first pointed out by Sen\cite{Send}.
Cosmologically, these defects will be produced\cite{GHY}.
Averaged over a region large compared to the typical defect
separation, this defect density becomes spatially uniform,
and may be identified as a part of the tachyonic matter.

In the brane world realization of superstring theory,
where the standard model particles are open string modes that live 
on the brane while the graviton and other closed string modes 
live in the bulk, the brane inflationary scenario\cite{dvali},
where the inflaton is simply an interbrane separation, 
looks very robust\cite{braneinf,rabadan,jst}.
In this scenario, tachyons
invariably emerge towards the end of brane inflation, when the 
branes approach each other and collide.
(Defects produced before or during inflation are inflated away.)
As the tachyon rolls down its potential, it was shown that
the only defects copiously produced are those that appear as
cosmic strings\cite{jst,costring}.
So we expect all of the tachyon potential energy goes to heating 
the Universe (via its coupling to the inflaton and other light
open string modes) and producing the cosmic string network.
Interactions of the cosmic strings will reduce the very high initial
cosmic string network density to an acceptable level\cite{vilenkin}, 
provided the cosmic string tension $\mu$ satisfies 
$G\mu <10^{-6}$, which seems to be always 
satisfied in brane inflation\cite{jst,costring}. 
The presence of cosmic strings will still give
detectable signatures in the cosmic microwave background and the
gravitational wave spectral density\cite{jst,costring}.
To summarize, the over-abundance of the tachyon matter density 
problem can be solved rather naturally by draining the tachyon 
potential energy to (re-)heating and to the cosmic string production.
Below, we discuss these two mechanisms.

Consider a general Lagrangian of the form ($\alpha^{\prime}=1$) 
\be
{\cal L}=-V(T)F(X)\qquad X\equiv\gupmunu\tmu\tnu~,
\label{laggen}
\ee
where $T(x)$ is the tachyon mode, $\tmu=\partial_{\mu}T$, and
$F(X)$ can be any function at this stage. 
In Ref.\cite{sen1}, $V(T) \propto e^{-T}$ and $F(X)=\sqrt{1+X}$.

The energy-momentum of Eq.(\ref{laggen}) is
\be
T_{\mu\nu}=V(T)\left[2\parX\tmu\tnu -\gmunu F(X)\right]~.
\label{tmunu}
\ee
where $F_{,X}=\partial F/\partial X$.
We can write this in the form of the energy-momentum tensor
for a perfect fluid if we identify
\ba
\rho_T &=& V(T)\left[F(X)-2X\parX\right]\equiv V(T)D(X) \nonumber \\
p_T &=& -V(T)F(X)~.
%\qquad U_\mu=-{\tmu\over\sqrt{-X}}
\label{prhou}
\ea
%and the equation of motion of the tachyon field is
%\be
%U^\mu\gradmu\rho+(\rho+p)\gradmu U^\mu=0~,
%\label{eqofmotion}
%\ee
%in the absence of dissipative processes.

We shall mostly use the tachyon Langrangian suggested by the 
boundary (or background-independent) string field theory\cite{bsft}, 
in particular,
its superstring version\cite{kutasov,superbsft}, which has been 
used to study tachyon matter recently \cite{sugimoto}.
The non-BPS brane model (with brane tension $\sqrt{2} \tau_p$) is,
\ba
V(T) &=& \sqrt{2} \tau_p e^{-T^2/2} \nonumber \\
 F(X) &=& {\sqrt{\pi}\Gamma(1+X)\over\Gamma(X+1/2)}
= {X4^X \Gamma(X)^2\over {2\Gamma(2X)}}~.
\label{bsfta}
\ea
%Here, $F(X) = 1 + X \ln4 $ for small $X$.
%In this model, 
For spatially independent solutions,  
$p_T<0$ for $\dot T^2<1/2$ and $p_T>0$ for $\dot T^2>1/2$.
%when $X=-\dot T^2$ is small and becomes positive when $X<-1/2$.
As $\dot T^2=-X\to 1$, $F(X)\simeq -1/2(1-\dot T^2)$,
%As $1+X= 1 - \dot T^2 \to 0$, $F(X) \simeq -1/{2(X+1)}$, 
$D(X) \simeq (1-\dot T^2)^{-2}$, so
%\be
$
w_T=p_T/\rho_T = -F(X)/D(X) \simeq (1-\dot T^2)/2 \to 0~.
$
%\ee
%As $1+X \to 0$, $V(T)\to 0$, but 
%the energy density $\rho_T$ stays constant, as required
%by energy conservation. 
The effective tachyon sound speed
$v_T=\sqrt{-F_{,X}/D_{,X}}\to 0$ for $\dot T^2\to 1$,
%also vanishes in this limit, 
allowing gravitational clustering. Since $w_T$ and
$v_T$ both vanish as $\dot T^2\to 1$, tachyon matter is
possible cold dark matter candidate, but substantial fine-tuning
is necessary for tachyon dark matter to be viable\cite{SW}.

Next, let us consider the brane-anti-brane pair, which has a complex
tachyon $T=T_1 + iT_2$.
The tachyon action is\cite{superbsft}
\be
{\cal S}_T=-2 \tau_p \int dx^{p+1} e^{- (T_1^2 +T_2^2)/2} F(X_1)F(X_2)
\label{baba}
\ee
where $X_I\equiv\gupmunu \partial_{\mu} T_I \partial_{\nu}T_I$.
Again, let us consider spatially independent solutions. 
The energy conservation requires that $\dot \rho_T=0$, where
\ba
\rho_T &=& 2 \tau_pe^{- \vert T\vert^2/2}\left[F_1F_2-2X_1F_{,X_1}F_2
 -2X_2F_1F_{,X_2}\right]
\ea
where $F_I=F(X_I)$ and  $F_{,XI}=F_{,X}(X_I)$.
First, consider a simple ansatz of linear tachyon profile
\be
T_1(t) = t - \epsilon(t)
\label{tacheps}
\ee
and $T_2=0$. 
One finds that 
\be
\dot \epsilon(t) \simeq e^{-t^2/4} \to 0
\label{epseq}
\ee
and 
$1+X_1 \to 0$, so $w_T=0$ (i.e., zero pressure). Since $X_2=0$ and
$F(X_2)=1$, this essentially reduces to the single $T$ case. In this 
solution, pressure approaches zero from the positive side\cite{sugimoto}.

Suppose we consider a more symmetric ansatz:
%\be
$T_1(t) = t - \epsilon_1(t)$, and $T_2(t)=t - \epsilon_2(t)$.
%\ee
An examination of the equations of motion yields  $\dot \epsilon_1 
\propto  \dot \epsilon_2 \propto e^{-t^2/3}$. However, when
these are substituted into $\rho_T$, we find
\be
\rho_T \simeq - \frac{\tau_p e^{-t^2}}{8}\left(\frac{1}
{\dot \epsilon_1^2\dot\epsilon_2}
+ \frac{1}{\dot \epsilon_1\dot \epsilon_2^2}\right)~.
\ee
A positive constant $\rho_T$ implies $\dot \epsilon_I<0$ (i.e., $1+X_I<0$),
%for at least one of the tachyon fields,
a solution that cannot be reached starting from small $\dot T_I$.

%The expansion rate of the Universe is simply
%%\be
%$H^2={8\pi G\rho/3}$.
%%\ee                     
%where $a(t)$ is the cosmological scale factor and
%$H=\dot a/a$. 
%For a spatially uniform tachyon field evolving in an expanding
%Universe, we have
%%Eq.(\ref{eqofmotion}) becomes
%%\be
%$\dot\rho=-3H\left(\rho+p\right)$.
%%\label{bckgndev}
%%\ee
%With $w=w_T=0$, 
%%Eq.(\ref{bckgndev}) implies 
%$\rho_T a^3={\rm constant}$, and since $v_T=0$ too,
%tachyon matter is dust-like\cite{sen1,sugimoto}.

Let us estimate the tachyon matter density today as a fraction $\Omega_T$
of the total density $3H_0^2M_{Pl}^2$ of our Universe.
We assume $\rho_{T,initial} \simeq M_s^4$, where the string scale 
$M_s^2= 1/\alpha^{\prime}$. Using $H_0^2/M_{Pl}^2 \simeq 10^{-123}$ and 
$a_{initial}/a_{today} \simeq 2.7^oK/M_s$, we find
\be
\Omega_T = \frac{\rho_{T,initial}}{3H_0^2M_{Pl}^2}
\left(\frac{a_{initial}}{a_{today}}\right)^3
\simeq 10^{28} \frac{M_s}{M_{Pl}}
\label{omega}
\ee
where $2.7^oK/M_{Pl} \simeq 10^{-31}$.
Since $\Omega_T$ must be less than 1, and $M_s >1$ TeV, we see that the 
tachyon matter density will be many orders of magnitude too big to be 
compatible with our Universe\cite{SW}. Fortunately, as we shall argue
below, this tachyon energy can be completely drained by cosmic string 
production and (re-)heating in a realistic early Universe.

One may also consider a stable solitonic solution of Eq.(\ref{baba}),
with $T_1(x)=u_1x_1$ and $T_2(x)=u_2x_2$\cite{superbsft}.
Since both the second derivatives of $T_I(x)$ and the mixed 
derivatives $\partial_iT_1\partial_jT_2$ with $i=j$ vanish 
in this ansatz, the equations of motion reduce to
$T_ID(X_I) = 0~(I=1,2)$.
Since $D(X_I)$ only vanishes at $X_I=u_I^2 \to \infty$
(this means infinitely thin brane),
the codimension-2 brane tension ($F(X) \to \sqrt{\pi X}$) becomes
\ba
\tau_{p-2} &=& 2 \tau_p \int dx_1 dx_2 e^{- \vert T\vert^2/2}
F(u_1^2)F(u_2^2) \nonumber \\
&=& 2 \tau_p F(u_1^2)F(u_2^2) 2\pi / {u_1 u_2} \to \tau_p (2 \pi)^2
\ea
as expected. This is a vortex solution.

A non-BPS $(p-1)$-brane may also be produced 
($u_1 \to 0, u_2 \to \infty$),
but it will quickly decay to BPS D$(p-2)$-branes.
%Multi-soliton solutions
%may also be constructed via the introduction of Chan-Paton factors.
A moving brane with constant velocity $v$ may also be constructed
by generalizing $T_1$ to $T_1(x)=u_1(x_1 -vt)$,
which gives the energy $E= \tau_{p-2}/\sqrt{1-v^2}$, as expected.
%Notice that the BSFT action is derived with a linear tachyon profile, so
%higher derivatives of $T$ are ignored in the above action. Fortunately,
%the soliton solution is given by linear profiles of $T$ only, where all
%second and higher derivatives of $T$ are exactly zero. So the
%solution is exact before the introduction of quantum corrections.
%Similarly, a stable soliton, i.e., a BPS D$(p-1)$-brane, emerges
%from the tachyon condensation for a non-BPS $p$-brane.

If we start with a brane-anti-brane pair, the tachyon will roll with
non-trivial spatial dependence. This will result in a collection of
D$(p-2)$-branes. However, if we average over a region much
larger than the typical spacing between the D$(p-2)$-branes, we expect 
to obtain a spatially uniform energy density and isotropized 
pressure, that is, a perfect fluid. In this sense, we may
interpret the density of defects (which turn out to be cosmic
strings \cite{jst,costring}) as part of the tachyon matter.
%In this sense, we may interpret the defect density as part 
%of the tachyon matter. 
%However, the fraction of energy that goes 
%to defect formation is sensitive to the details 
%and difficult to calculate.
%
%The cosmological production of defects in superstring theory has 
%been studied\cite{GHY}.
%In particular, D3-brane-anti-D3-brane annihilation yields 
%D1-branes\cite{Send}, which appear as cosmic strings in our Universe. 
%Actually, following from the fact that vortices (instead of 
%domain walls or monopoles) are formed, the appearance of only cosmic 
%strings is much more general\cite{jst,costring}, 
%As an illustration, consider D5-branes with two of its dimensions 
%compactified on a two-cycle, while the remaining 3 dimensions 
%span our Universe. D5-brane-anti-D5-brane annihilation will produce 
%D3-branes, which will also wrap around the two-cycle, so they also 
%appear as cosmic strings. If there is no one-cycle inside 
%the two-cycle, (e.g., topology of a sphere, as in orbifolding a torus), 
%then no domain wall or monopole-like solitons will appear in our Universe.
%
The Kibble mechanism would imply an initial density $\sim H^2M_s^2
\sim M_s^6/M_{Pl}^2$ in cosmic strings at the end of brane inflation.
Intercommutation of intersecting cosmic stings and the decay of
string loops (to gravitational radiation) causes the density of
the cosmic string network (CSN) to approach $\Omega_{CSN}\sim G\mu\sim
M_s^2/M_{Pl}^2$ \cite{vilenkin}, which would be acceptable.
%$\sim M_s^2T^4/M_{Pl}^2$ at temperature $T$, which would be acceptable.
However, the initial density of the network is only a fraction
$\sim M_s^2/M_{Pl}^2$ of the total density of the Universe,
$\sim M_s^4$, at the
end of brane inflation. This starting density must be overwhelmingly in
the form of radiation in order for the cosmology to be viable. Thus,
we must understand how the mass density in tachyons can convert to
radiation efficiently at the end of brane inflation.

%Naively, the cosmic string network has a density that is comparable 
%to and decreases like (for cosmic string loops) that of the tachyon 
%matter (\ref{omega}), or, for cosmic strings stretching across the 
%horizon, even slower (i.e., worse). 
%Fortunately, the intercommutation of intersecting cosmic strings and the 
%decay of cosmic string loops (to gravitational waves) modify the 
%density to decrease like radiation\cite{vilenkin}. 
%This adds another power of $2.7^oK/M_s$ to
%Eq.(\ref{omega}). 
%rendering the cosmic string network density to 
%an acceptable level.

To drain the tachyon energy to heat the Universe, we need to couple 
the tachyon field to other fields, the most obvious one being the scalar 
field $\phi$ that describes the separation between the 
brane and the anti-brane ($\phi=0$ in the above discussion). 
In brane inflation, $\phi$ is the inflaton.
%Consider what happens in a model where the
%tachyon appears during the last stages of the
%inflation era, is present during heating, and also afterwards. 
Let us take a simple generalization of the above
tachyon model to include the inflaton $\phi$,
\be
{\cal L}(T,\phi) =-V(\vert T\vert,\phi)F(X_1)F(X_2)
- {1\over 2}g^{\mu\nu}\phimu\phinu~.
\label{actionph}
\ee
For brane inflation, the inflaton potential is known to have 
the form $U(\phi) \simeq U_0-{U_1/ \phi^{d-2}}$
for large $\phi$, when there are $d$ extra
dimensions transverse to the brane\cite{braneinf,gia}. 
For a particular brane pair,
$U_i$ are easily calculable \cite{polchinski}. For example, for
a brane-anti-brane pair, $U_0=2 \tau_pV_{p-3}$ where
$V_{p-3}$ is the compactification volume of the
extra $(p-3)$ dimensions of the branes and $U_1$ is simply twice 
that of the gravitational force. 
%(Note that only the lightest
%Kaluza-Klein modes of both $\phi$ and $T$ are kept in the above
%effective theory.) 
When the brane separation $\phi$ is 
comparable to the compactification size, the compactification 
effect becomes important and it flattens the potential 
further\cite{braneinf,jst}, allowing inflation to take place for 
many e-foldings. 

When the branes are far apart, $T$ is a normal scalar field with
positive mass squared $m_T^2$. As the branes move closer, $m_T^2$ decreases, 
and becomes negative at the bifurcation point $\phi_v$, where 
$T$ becomes tachyonic. This suggests the following potential:
\ba
V(\vert T\vert,\phi) &=& U(\phi)
\exp\left[-{\vert T\vert^2}\left({1\over 2}-\lambda \phi^2\right)\right] 
\label{tainf}
\ea
where the bifurcation point $\lambda \phi_v^2=1/2$. 

Although the potential $U(\phi)$ for small $\phi$ remains to be
calculated, it is expected to be positive and finite as $\phi \to 0$. 
In terms of open strings, $U(\phi=0)$ is a quantum 
one-loop correction, so $2\tau_p$ is renormalized to $2 \tau_pv_0$, 
where we shall take $1 > v_0 > 0$.
So, for $\phi=0$, we recover the tachyonic model (\ref{baba}). 
Notice that the minimum of the potential $V(\vert T\vert,\phi)$
($\phi<\phi_v$ and $T \to \infty$) is automatically zero, 
independent of the details of $U(\phi)$. As a consequence,
the dynamics is insensitive to the particular form of $U(\phi)$ 
for $\phi<\phi_v$, since tachyon rolling 
happens very fast, and $V(\vert T\vert,\phi) \to 0$ rapidly.
For phenomenology, we choose a particularly simple form of $U(\phi)$,
where $U(\phi)$ has a simple harmonic form for small $\phi$,
\be
U(\phi)=U_0\left[1-{(1-v_0)\over(\phi^2/\phi_0^2+1)^{{1\over 2}(d-2)}}
\right]~.
\label{infpot}
\ee
In summary, Eqs.(\ref{actionph},\ref{tainf},\ref{infpot})
comprise the model, which may be considered as a novel version 
of hybrid inflation. 

For fixed $\phi >\phi_v$, $T$ has positive mass $m_T^2 >0$, 
so the minimum of the potential is at $T_I=0$ and classically 
$\partial_{\mu} T_I=0$, so $X_I=0$
and $F(X_I)=1$. The model reduces to
${\cal L} (\phi, T=0) =- U(\phi)
- g^{\mu\nu} \partial_{\mu} \phi\partial_{\nu} \phi/2$.
This describes the inflationary epoch for $\phi>\phi_v$.

Based on the above analysis, let only one of the tachyons roll,
say $T=T_1$, while $T_2$ is essentially frozen, with $F(X_2)=1$. 
The analysis applies just as well if we replace $F(X_1)F(X_2)$
in Eq.~(\ref{actionph}) with the U(1) invariant $F(X_1+X_2)$.
For a spatially uniform scalar field and tachyon
in a FRW expanding Universe,
\ba
\ddot\phi+(3H+\Gamma_\phi)\dot\phi&=&
-F(X)[V(\vert T\vert,\phi)]_{,\phi}\nonumber\\
\ddot T+3HA(X)\dot T&=&-{V_{,T}\over 2V}B(X)-{V_{,\phi}\over V}
A(X)\dot T\dot\phi~,
\label{cosmoeqns}
\ea
where $A(X)=F_{,X}/(F_{,X}+2XF_{,XX})$ and 
$B(X)=(F-2XF_{,X})/(F_{,X}+2XF_{,XX})$; $\Gamma_\phi$
is a phenomenological term that models the decay of 
$\phi$ to ordinary particles and radiation, whose density is
$\rho_{matter}$. If the inflaton decays primarily to
relativistic particles (``radiation''), then
\be
\dot\rho_{matter}+4H\rho_{matter}=\Gamma_\phi\dot\phi^2~.
\label{rhomatter}
\ee
The total energy density in this situation is
\be
\rho=V(\vert T\vert,\phi)(F(X)+2\dot T^2\FX)
+ \frac{\dot\phi^2}{2} + \rho_{matter}
\ee
and $H^2=8\pi G\rho/3$.
Here, we are not interested in the entire process of
brane inflation, just what happens at the end,
when the tachyon starts to roll. We assume that there
is a prior period of slow rolling ($\phi$ decreasing slowly), 
during which almost
all of the inflation occurs, and density fluctuations
are generated\cite{braneinf,rabadan,jst}.
During the inflationary period, when $\vert T\vert\phi$ 
is small and (let $\lambda=1$) $\phi>2^{-1/2}$, the evolution 
of $T(t)$ resembles a massive field, with a mass $\sim\phi$, 
and we therefore expect $T(t)$ to remain small, and
dominated by quantum fluctuations. When the tachyon
becomes unstable, it begins with an amplitude 
$T\sim H_I/2\pi\sqrt{U_0}\sim M_{Pl}^{-1}$ 
and $\dot T\sim H_IT$, where $H_I=(8\pi GU_0/3)^{1/2}$
is the expansion rate during slow roll. We follow the
evolution implied by Eq. (\ref{cosmoeqns}) and
(\ref{rhomatter}) after $\phi=2^{-1/2}$. 

%
%\begin{figure}[h]
\newcounter{figcount}
\setcounter{figcount}{1}
\begin{center}
\epsfxsize=3.5in
\epsfbox{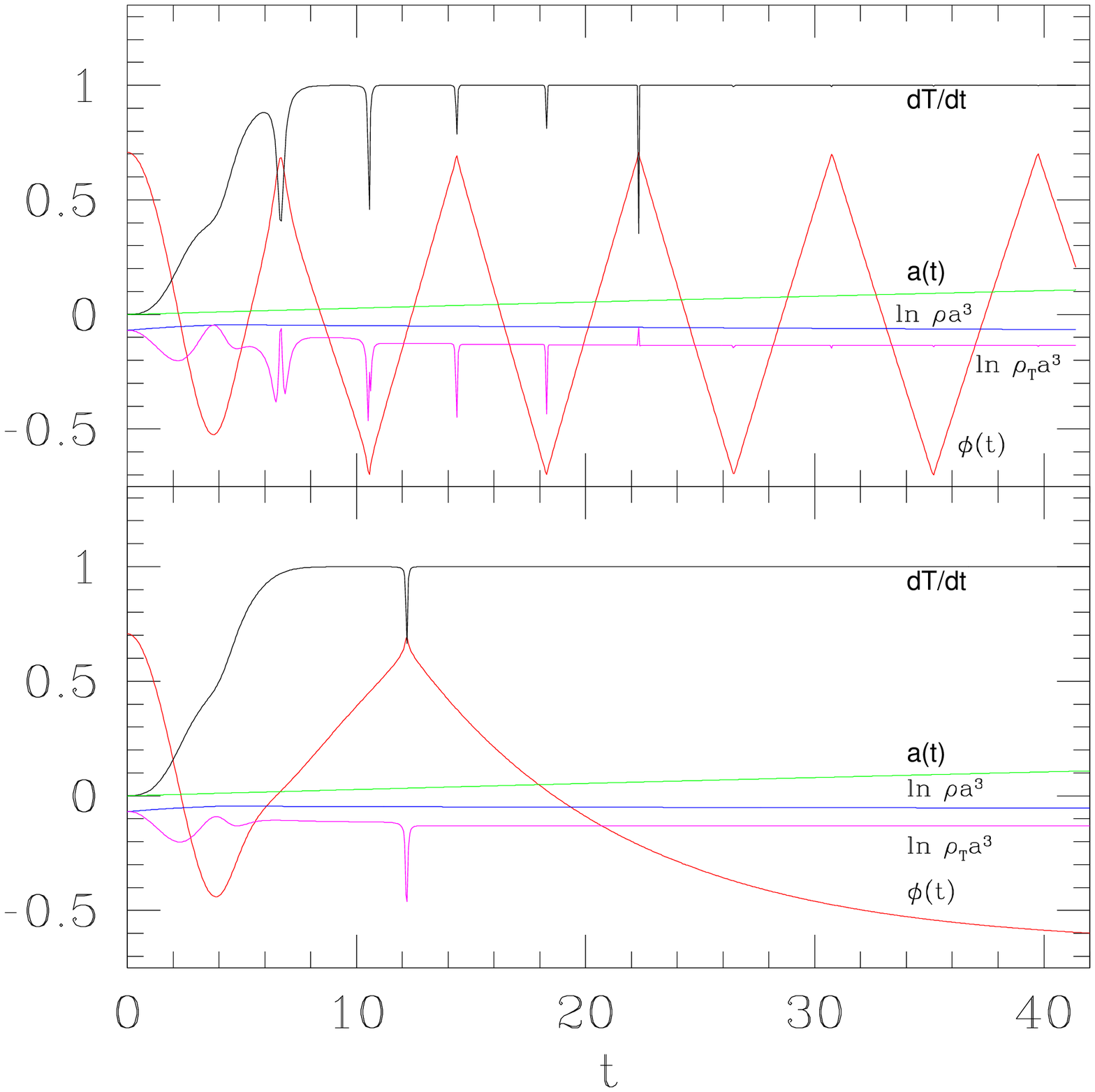}
%\caption{
\parbox{8cm}{\vspace{12pt}
FIG. \thefigcount\hspace{12pt}
Cosmological evolution of
$\phi(t)$ (red), $\dot T$ (black), $\ln a(t)$
(green), $\ln\rho a^3$ (blue) and $\ln\rho_T
a^3$ (magenta) for $\Gamma_\phi=0$ (top)
and $\Gamma_\phi=0.1$ (bottom). $M_{Pl}=10^3$,
$v_0=0.9$, $U_0=1$.
\vspace{12pt}
}
\end{center}
%\end{figure}

As an illustration, numerical results are shown in Fig. 1, 
assuming (in terms of string scale) $1/M_{Pl}=10^{-3}$,
$U_0=1$, 
$1-v_0=0.1$, $d=4$, $\phi_0=1$, and $\Gamma_\phi=0$
or $0.1$. In fact, because $H_I\simeq 0.003$, cosmological
expansion is relatively unimportant. The simulations start
with $\phi(t)=1/\sqrt{2}$ and $\dot\phi=0$, at $t\equiv 0$
and $a(t)\equiv 1$. The initial data can be varied, but the
behavior shown in Fig. 1 is generic. At first, $\phi(t)$
decreases, and $\dot T$ increases. Ultimately, as
$\dot T\to 1$, $T\simeq t$ grows, and the product
$V(T,\phi)F(X)$ decreases rapidly for $\phi^2<1/2$.
In the nondissipative case, $\Gamma_\phi=0$, $\dot\phi
a^3\to$ constant, and since $a(t)$ is only changing
slowly, $\phi$ increases or decreases linearly with
time. However, since the total energy is
conserved, and $V(T,\phi)$ would grow exponentially
large for $\phi^2>1/2$, the growth of $\vert\phi\vert$
cannot continue indefinitely, so $\phi(t)$ bounces
near $\phi^2=1/2$. Associated with these bounces are
short-lived dips in $\dot T$, which are evident but
not completely resolved in Fig. 1. Eventually, cosmological
expansion alone would cause $\dot\phi$ to decay, and
$\phi$ would settle to some constant, nonzero value.
Dissipation accelerates the decay of the nonlinear
oscillations. The bottom panel in Fig. 1 shows what
happens for $\Gamma_\phi=0.1$, which may be unrealistically
large, but illustrates how dissipation suppresses the
oscillations, and causes $\phi$ to settle to a nonzero
value asymptotically. For either $\Gamma_\phi$,
$\rho_Ta^3\to$ constant $\sim 1$ eventually, where
$\rho_T\equiv V(T,\phi)D(X)$.
For $\Gamma_\phi=0.1$, $\rho_T/\rho\gtrsim 0.9$ at the end of
the numerical integration; the Universe
does not become radiation dominated.

%%
%%\begin{figure}[h]
%\newcounter{figcount}
%\setcounter{figcount}{1}
%%\setcounter{conspapereps}{\value{figcount}}
%\begin{center}
%\epsfxsize=3.5in
%\epsfbox{tachphiFpaper.ps}
%%\caption{
%\parbox{8cm}{\vspace{12pt}
%FIG. \thefigcount\hspace{12pt}
%Cosmological evolution of
%$\phi(t)$ (red), $\dot T$ (black), $\ln a(t)$ 
%(green), $\ln\rho a^3$ (blue) and $\ln\rho_T
%a^3$ (magenta) for $\Gamma_\phi=0$ (top)
%and $\Gamma_\phi=0.1$ (bottom). $M_{Pl}=10^3$,
%$v_0=0.9$, $U_0=1$.
%\vspace{12pt}
%}
%\end{center}
%%\end{figure}

Radiation domination can be achieved if the tachyon
couples directly to matter. A possible mechanism might 
be tachyon preheating \cite{kofman}. To explore a
different possibility phenomenologically, 
suppose we add a dissipative term $-\Gamma_T\dot T$
to the second of Eqs. (\ref{cosmoeqns}), and assume
that the energy loss due to this term goes directly
to matter. Then the rate of increase of the matter
density due to this form of coupling is
$2\Gamma_TXD_{,X}V(T,\phi)$. Assuming that the 
inflaton, $\phi$, also couples to matter as in the
above model, we find that eventually $\phi\to
\phi_\infty$, with $\vert\phi_\infty\vert<1/\sqrt{2}$,
while $\dot\phi\to 0$.
The tachyon rolls dissipatively toward its minimum
at late times, and balancing the drag 
$-\Gamma_T\dot T$ with the accleration down the
tachyon potential, $-V_{,T}B(X)/2V$, implies that
\be
1-\dot T\simeq
{2\Gamma_T\over t(1-2\phi_\infty^2)}~.
\label{dragone}
\ee
Thus, $1+X\simeq 2(1-\dot T)\to 0$ asymptotically, but only as
a power of time, whereas $V(T,\phi)\simeq
U(\phi_\infty)\exp[-t^2(1/2-\phi_\infty^2)]$ drops
exponentially; the rate of decrease of $1+X$ 
is far slower than in
Eqs. (\ref{tacheps}) and (\ref{epseq}).
After a time of order the maximum
of $\Gamma_T^{-1}$ and $\Gamma_\phi^{-1}$, the 
Universe becomes radiation dominated, and the tachyon
matter density is exponentially small.
Depending on $\Gamma_T$ and $\Gamma_\phi$,
there could be an early phase of tachyon domination,
but it would not last long enough to have any
important, lasting effects. Although
tachyon decay to matter never ceases in this model,
the radiation production rate
becomes exponentially
small at late times, and the additional heating 
long after the end of inflation is inconsequential.

It is possible that $\Gamma_T$ is a function of $T$, 
not constant. Eq. (\ref{dragone}) already gives some idea
of what might happen. If $\Gamma_T$ decreases with 
increasing $T$, then we expect that eventually dissipation
becomes relatively unimportant, and the Universe may
be left with a residual tachyon matter density. On the
other hand if $\Gamma_T$ increases with increasing $T$,
then we might expect $\dot T$ to fall toward zero eventually,
with $T$ asymptoting to some constant value or else growing
very slowly.

Let us consider first what happens when $\Gamma_T$ is
a decreasing function of $T$. Suppose, as is plausible,
that $\Gamma_\phi$ is characterized by the string scale,
so that $\phi$ tends toward a time-independent value
as branes approach one another relatively quickly,
perhaps even faster than expansion. For $H\sim M_s^2/
M_{Pl}$, and $\Gamma_\phi=\beta M_s$, where $\beta$
is dimensionless, $\Gamma_\phi/H\sim \beta M_{Pl}/M_s$,
which is greater than one as long as $\beta\gtrsim
M_s/M_{Pl}$. As $\dot\phi\to 0$ and $\dot T\to 1$, the
evolution equation for $\rho_T$ simplifies to
\be
\dot\rho_T+3H\rho_T\simeq -{V(T)\Gamma_T(T)\over
2(1-\dot T)^3}\simeq -{4\Gamma_T(T)\rho_T^{3/2}\over
\sqrt{V(T)}}~,
\label{rhoteqn}
\ee
where we used $\rho_T\simeq V(T)/4(1-\dot T)^2$ to
obtain the last form of the RHS. Eq. (\ref{rhoteqn})
has the solution
\be
\rho_Ta^3\simeq\left[u_\phi
+2\int_{t_\phi}^t{dt^\prime
~\Gamma_T(t^\prime)\over[a(t^\prime)]^{3/2}
\sqrt{V(t^\prime)}}\right]^{-2}~,
\label{rhotsol}
\ee
where $u_\phi$ is the value of
$(\rho_Ta^3)^{-1/2}$ at roughly the time $t_\phi$ when $\phi$
freezes; assuming a density $\sim M_s^4$ and scale factor
$\sim T_0/M_s$, where $T_0$ is the present temperature of
the Universe, $u_\phi\sim (M_s/T_0)^{3/2}\sim
10^{47.6}(M_s/M_{Pl})^{3/2}$.
In obtaining Eq. (\ref{rhotsol}), we have
approximated $T(t)\simeq t$ in the integrand. Eq.
(\ref{dragone}) can be recovered from Eq. (\ref{rhotsol})
by evaluating the integral asymptotically under the
approximation that $\Gamma_T$ is independent of $T$,
and expansion is slow compared to the string timescale.
For constant $\Gamma_T$, the tachyon matter density
falls rapidly with time, roughly like 
$\exp[-t^2(1/2-\phi_\infty^2)]$, and becomes utterly
negligible. As long as $\Gamma_T$ does not fall faster
than $\sqrt{V(T)}$, then the integral in Eq.
(\ref{rhotsol}) grows very rapidly.
In general, Eq. (\ref{rhotsol}) implies that 
for an acceptable $\rho_T$ today
\ba
\int_{t_\phi}^{t_0}{dt^\prime~\Gamma_T(t^\prime)\over
[a(t^\prime)]^{3/2} 
\sqrt{V(t^\prime)}}&\gtrsim 10^{14}\sqrt{M_s\over\Omega_TM_{PL}}u_\phi
\nonumber\\&\sim 10^{61.6}\Omega_T^{-1/2}\left({M_s\over M_{Pl}}\right)^{2}~,
\label{rholimsol}
\ea
if $\Gamma_T$ is a decreasing function of $T$; 
$t_0\gg t_\phi$ is the present age of the Universe. 
Although it is possible for tachyon matter to become
dominant in models with decreasing $\Gamma_T$, 
extreme  fine-tuning is required in order for
tachyons to be the cold dark matter. Since $1/\sqrt{V(t)}$
increases so rapidly -- like a Gaussian in $t$ --
decay of the tachyon field would have
to shut off completely after only $\sim 10M_s^{-1}$
for Eq. (\ref{rholimsol}) to allow substantial
$\rho_T$ today. Otherwise, the value of $\rho_T$ today
ought to be utterly negligible. Generically, we expect that
$\Gamma_T$ cannot shut off effectively and quickly enough
to leave a substantial tachyon matter density today.

It is also conceivable that $\Gamma_T$ depends on both
$T$ and $\dot T$. As $\dot T\to 1$, suppose that
$\Gamma_T=(1-\dot T)^p\gamma_T(T)$. Then instead of
Eq. (\ref{rhotsol}) we would find
\be
\rho_Ta^3\simeq\biggl[u_\phi^{1-p}
+2^{1-p}
\int_{t_\phi}^t{dt^\prime
~(1-p)\gamma_T(t^\prime)\over[a^3(t^\prime)
V(t^\prime)]^{{(1-p)\over 2}}}\biggr]^{-{2\over (1-p)}}~.
\label{rhotpsol}
\ee
For $p<1$, Eq. (\ref{rhotpsol}) is similar to Eq. 
(\ref{rholimsol}), and implies a leftover tachyon matter
density at late times which may or may not be important
depending on the details of the damping rate. 
For $p>1$, the integrand in Eq. (\ref{rhotpsol}) is
proportional to $[V(t)]^{(p-1)/2}$, which falls like
a Gaussian, so we expect relatively little decrease in
$\rho_Ta^3$, for $\gamma_T\lesssim 1$. The special
value $p=1$ corresponds to a dissipation rate proportional
to $\rho_T$, and in that case we find
\be
\rho_Ta^3\simeq(\rho_Ta^3)_\phi\exp\left[-2\int_{t_\phi}^t
{dt^\prime\gamma_T(t^\prime)}\right]~.
\ee
The final value of $\rho_Ta^3$ depends sensitively on how
quickly $\gamma_T(T)$ decreases with increasing $T$ for
$p=1$.

Next, let us briefly consider the opposite case,
where $\Gamma_T$ increases with $T$ faster than
$T$. In that case, $\dot T$ eventually drops 
toward zero, and $T(t)$ changes only slowly. 
If the dissipative drag approximately balances
acceleration down the tachyon potential then once
$\phi$ has frozen
\be
\Gamma_T\dot T\simeq {T(1-2\phi_\infty^2)\over
2\ln 4}~,
\ee
Small $\dot T$ requires small $T/\Gamma_T$. 
For this regime to extend to small $\rho_T\simeq
V(T)$, we would have to require that $\Gamma_T/T$
is large even for moderately large values of $T$. But this would
also imply that $\Gamma_T$ becomes large in units
of the string scale, which would not be expected
to occur in perturbation theory; in a nonperturbative
calculation, back reaction ought to prevent the growth
of $\Gamma_T$ to large values. Thus, if $\Gamma_T$
increases at all, it must only do so for a short time,
lasting only $\sim M_s^{-1}$.
Thereafter, $\Gamma_T$ could level 
off to a constant value or decrease, and the
physics reverts to Eq. (\ref{rhotsol}) (or Eq.
[\ref{dragone}]).

Possible products of tachyon decay could include ordinary
elementary particles, gravitational radiation and
massive defects. Of these, ordinary elementary
particles would be the most benign, since they would
simply thermalize to reheat the Universe. Since most
of the decrease in tachyon energy density must occur
within $\sim 10M_s^{-1}$ of brane collision, the 
wavelengths of any gravitons produced are expected
to be small, at most of order $\sim\zeta M_{Pl}/M_s
T_0\sim 0.1\zeta M_{Pl}/M_s$ cm, where $\zeta\sim 
10$ or so. This estimate presumes that the Universe
becomes radiation dominated -- maybe by gravitational
radiation predominantly -- at a density $\sim (M_s/\zeta)^4$, and 
that gravitational radiation produced at this epoch
has wavelengths smaller than the horizon scale
$\sim\zeta^2M_{Pl}/M_s^2$, which redshifts by a factor
of $\sim M_s/\zeta T_0$ to the present day. The most stringent 
limit on the density of
short wavelength gravitational radiation comes from
the requirement that the energy density in gravitons
is sufficiently small at the time of nucleosynthesis.
Suppose that the end result of reheating is a fraction
$f_g$ in gravitons, with the rest in ordinary elementary
particles (perhaps produced only by the decay of the
inflaton). If the effective number of spin degrees
of freedom of relativistic particles in equilibrium
after reheating is $S_{rh}$, and $S_{nuc}$ is
the analogous value at nucleosynthesis, then we
expect that the fraction of the energy density in
gravitons at nucleosynthesis is $\simeq f_g
(S_{nuc}/S_{rh})^{4/3}$. ($S_{nuc}=43/8$ before
neutrinos go out of equilibrium at $T\simeq 1$
MeV, $S_{nuc}=11/4$ between 1 MeV and 
$e^\pm$ annihilation at $\sim m_e$,
and $S_{nuc}=1$ after $e^\pm$ annihilation.) 
Assuming that gravitons can only comprise at most
$\epsilon_{nuc}$ of the total energy density of the Universe
at the nucleosynthesis epoch implies that 
$f_g(S_{nuc}/S_{rh})^{4/3}\lesssim\epsilon_{nuc}$;
for $\epsilon_{nuc}\sim 0.01-0.1$, this limit can be
satisfied as long as $S_{rh}/S_{nuc}\gtrsim 10-100$,
which is not implausibly large. Massive 
defects that decay to elementary particles or
gravitons could also be acceptable, provided that
they decay early enough.

This model does not account for any small scale
inhomogeneities in $T({\bf x},t)$, and so does not
describe formation of cosmic strings. 
The decay of the tachyon energy density is not
quite as efficient as outlined above because cosmic strings
form via the Kibble mechanism. However, the density
of cosmic strings forming in this way is expected
to be small, and, at least within the context of
the $\Gamma_T\neq 0$ model described above, could
be far smaller than the density in
radiation at the end of brane inflation, and small
enough not to be ruled out by currently available
cosmological observations. 

To summarize, we have shown that the tachyon matter
density problem can be solved via dissipation and
reheating. Generically, such mechanisms, if active,
ought to result in a negligible tachyon matter density
today; extreme fine-tuning is needed for tachyon matter
to be the cold dark matter.
Eq. (\ref{rholimsol}) gives a numerical
constraint that must be satisfied by such a mechanism.
Whether or not there are realistic physical scenarios
to achieve the required tachyon decay remains to be seen.

We thank J. Cline, N. Jones,
L. Kofman, A. Linde, A. Naqvi, A. Sen and S. Shatashvili 
for discussions.
This research is partially supported by the DOE grant EY-76-02-3071 
(G.S.), UPenn SAS Dean's fund (G.S.), NSF (S.-H.H.T.), and NASA
(I.W.).


\begin{references}

\bibitem{sen1} A.~Sen,~hep-th/0203211;~hep-th/0203265;~hep-th/0204143.

\bibitem{gibbons} G.W.~Gibbons, hep-th/0204008;
A.~Frolov, L.~Kofman and A.~Starobinsky, hep-th/0204187;
L.~Kofman and A.~Linde, hep-th/0205121.

\bibitem{SW} G.~Shiu and I.~Wasserman, hep-th/0205003.

\bibitem{Send} A.~Sen, JHEP {\bf 9808} (1998) 012, hep-th/9805170;
E.~Witten, JHEP {\bf 9812} (1998) 019, hep-th/9810188;
P.~Horava, Adv. Theor. Math. Phys. {\bf 2} (1999) 1373, hep-th/9812135.

\bibitem{GHY}G.W.~Gibbons, K.~Hori and P.~Yi, Nucl. Phys. {\bf B596} 
(2001) 136; M.~Majumdar and A.-C.~Davis, hep-th/0202148;
K. Hashimoto, hep-th/0204203.

\bibitem{dvali}
G.~R.~Dvali and S.-H.H.~Tye,
Phys. Lett. {\bf B450}, (1999) 72, hep-ph/9812483.

\bibitem{braneinf}
C.P.~Burgess, M.~Majumdar, D.~Nolte, F.~Quevedo,
G.~Rajesh and R.J.~Zhang, JHEP {\bf 0107} (2001) 047, hep-th/0105204.

\bibitem{rabadan} J.~Garcia-Bellido, R.~Rabadan and F.~Zamora, 
JHEP {\bf 0201} (2002) 036, hep-th/0112147;
C.~Herdeiro, S.~Hirano and R.~Kallosh,
JHEP {\bf 0112}, (2001) 027,
hep-th/0110271.
B.~S.~Kyae and Q.~Shafi,
Phys. Lett. {\bf B526} (2002) 379,
hep-ph/0111101.

\bibitem{jst}N.~Jones, H.~Stoica and S.-H.H.~Tye,
hep-th/0203163.

\bibitem{costring}S.~Sarangi and S.-H.H.~Tye, hep-th/0204074.

\bibitem{vilenkin} A.~Albrecht and N.~Turok, Phy. Rev. Lett. {\bf 54}
(1985) 1868; D.P.~Bennett and F.R.~Bouchet, ibid, {\bf 60} (1988) 257;
B.~Allen and E.P.S.~Shellard, ibid, {\bf 64} (1990) 119.

\bibitem{bsft} E.~Witten, Phys. Rev. {\bf D46} (1992) 5467, 
hep-th/9208027; ibid, {\bf D47} (1993) 3405, hep-th/9210065;
S.~Shatashvili, Phys. Lett. {\bf B311} (1993) 83, hep-th/9303143.

\bibitem{kutasov} D.~Kutasov, M.~Marino amd G.W.~Moore, hep-th/0010108.

\bibitem{superbsft} P.~Kraus and F.~Larsen, hep-th/0012198;
T.~Takayanagi, S.~Terashima and T.~Uesugi, hep-th/0012210.

\bibitem{sugimoto}
S.~Sugimoto and S.~Terashima, hep-th/0205085;
J.A.~Minahan, hep-th/0205098.

\bibitem{gia}G.~Dvali, Q.~Shafi and S.~Solganik, hep-th/0105203. 

\bibitem{polchinski} See e.g., J. Polchinski, {\it String Theory},
Cambridge University Press (1998).

\bibitem{kofman} G.~Felder, J.~Garcia-Bellido, P.B.~Greene, L.~Kofman,
A.~Linde and I.Tkachev, Phy. Rev. Lett. {\bf 87} (2001) 011601; 
G.~Felder, L.~Kofman and A.~Linde, hep-th/0106179.

\end{references}
\end{document}